\documentclass[aps,prb,article,twocolumn,preprintnumbers,amsmath,amssymb,superscriptaddress, longbibliography]{revtex4-1}

\date{\today}
\usepackage{epsfig}
\usepackage{cases}
\usepackage{subfigure}
\usepackage{amsmath}
\usepackage{graphicx}
\usepackage{dcolumn}
\usepackage{bm}
\usepackage[colorlinks,linkcolor=blue,hyperindex,CJKbookmarks]{hyperref}
\usepackage{float}
\usepackage{mathtools}
\usepackage{hyperref}
\usepackage{comment}
\usepackage{bbm}
\begin{document}
\newcommand{\Ham}{\mathcal{H}}
\newcommand{\kbf}{\mathbf{k}}
\newcommand{\qbf}{\mathbf{q}}
\newcommand{\Qbf}      {\textbf{Q}}
\newcommand{\lbf}      {\textbf{l}}
\newcommand{\ibf}      {\textbf{i}}
\newcommand{\jbf}      {\textbf{j}}
\newcommand{\rbf}      {\textbf{r}}
\newcommand{\Rbf}      {\textbf{R}}
\newcommand{\Schrdg} {{Schr\"{o}dinger}}
\newcommand{\aband} {{(\alpha)}}
\newcommand{\bband} {{(\beta)}}

\newcommand{\eps} {{\bm{\varepsilon}}}
\newcommand{\probA}      {{\mathsf{A}}}
\newcommand{\HamPump}      {{\Ham_{\rm pump}}}
\newcommand{\HamPr}      {{\Ham_{\rm probe}}}
\newcommand{\timeMax} {{t_{\rm m}}}
\newcommand{\qin} {{\qbf_{\rm i}}}
\newcommand{\qout} {{\qbf_{\rm s}}}
\newcommand{\epsin} {{\eps_{\rm i}}}
\newcommand{\epsout} {{\eps_{\rm s}}}
\newcommand{\win} {{\omega_{\rm in}}}
\newcommand{\wout} {{\omega_{\rm s}}}

\title{Probing light-driven quantum materials with ultrafast resonant inelastic X-ray scattering}
\author{Matteo Mitrano}
\affiliation{Department of Physics, Harvard University, Cambridge, Massachusetts 02138, USA}
\email{mmitrano@g.harvard.edu}
\author{Yao Wang}
\affiliation{Department of Physics and Astronomy, Clemson University, Clemson, South Carolina 29631, USA}
\email{yaowang@g.clemson.edu}
\date{\today}

\begin{abstract}
\begin{center}
\textbf{\abstractname}
\end{center}
Ultrafast optical pulses are an increasingly important tool for controlling quantum materials and triggering novel photo-induced phase transitions. Understanding these dynamic phenomena requires a probe sensitive to spin, charge, and orbital degrees of freedom. Time-resolved resonant inelastic X-ray scattering (trRIXS) is an emerging spectroscopic method, which responds to this need by providing unprecedented access to the finite-momentum fluctuation spectrum of photoexcited solids. 
In this Perspective, we briefly review state-of-the-art trRIXS experiments on condensed matter systems, as well as recent theoretical advances. We then describe future research opportunities in the context of light control of quantum matter.
\end{abstract}
\maketitle

\section{Introduction}
Understanding and controlling quantum materials --- material systems exhibiting quantum-mechanical effects over wide energy and length scales\cite{Keimer2017} --- is a central challenge in modern condensed matter physics. Over the last two decades, ultrafast lasers have had a tremendous impact on quantum materials research and provided a novel route to on-demand engineering of their electronic and structural properties. They have not only allowed for tuning well-known states of matter far from equilibrium, e.g. magnetism \cite{Stanciu2007,Nova2016,Afanasiev2019,Hendry2019,Disa2020}, charge/spin order \cite{Schmitt2008,perfetti2008femtosecond,Rohwer2011,Kim2012,Lee2012}, and ferroelectricity \cite{kubacka2014,Nova2019,Li2019}, but also led to novel dynamical phenomena, such as transient superconductivity \cite{Fausti2011,Hu2014,Kaiser2014,Mitrano2016} and Floquet topological phases \cite{Wang2013,Mahmood2016,McIver2019}. 

\begin{figure}
    \centering
    \includegraphics[width=\columnwidth]{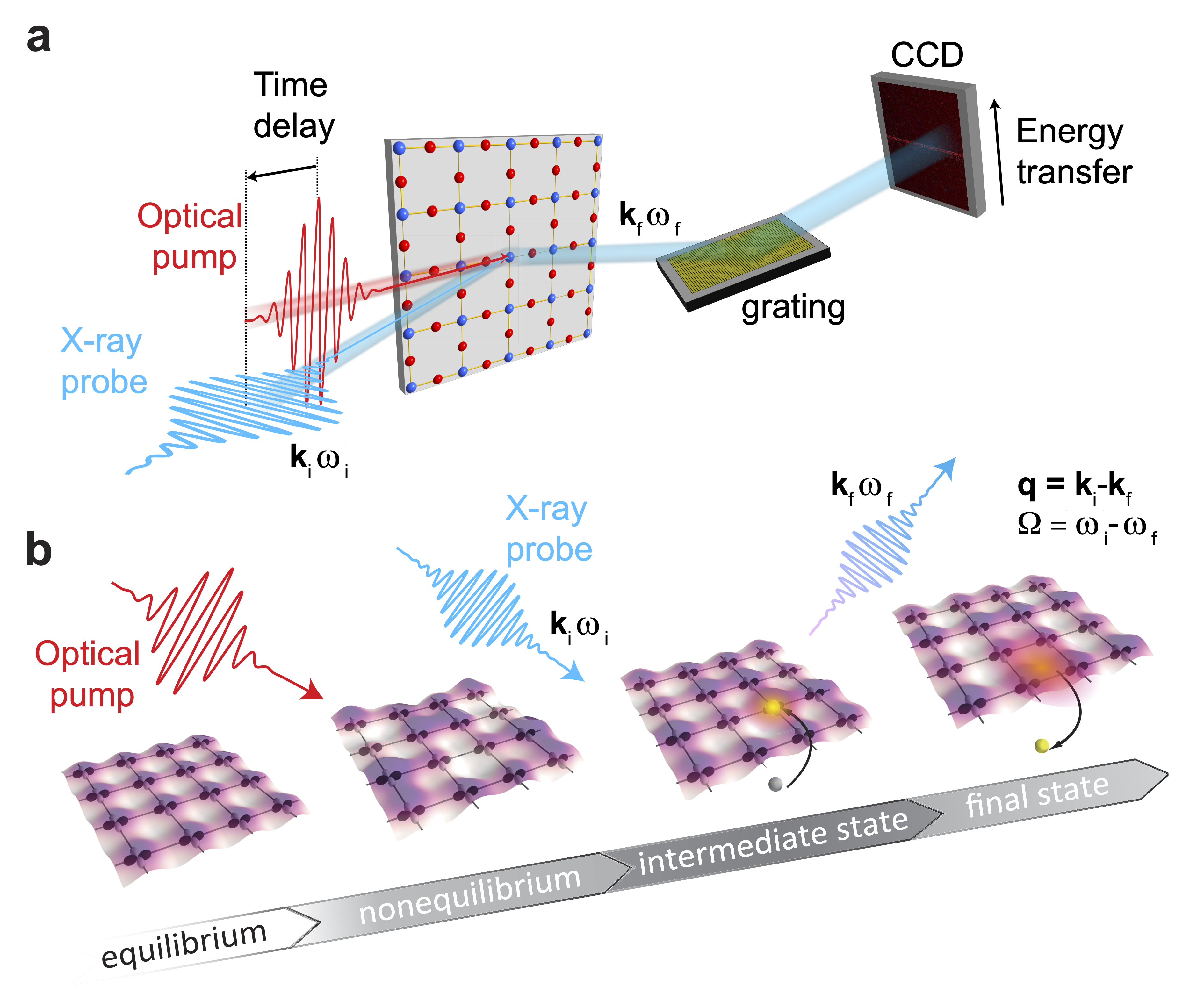}
    \caption{ \textbf{Basics of a time-resolved resonant inelastic X-ray scattering (trRIXS) experiment.} {\bf a} A material is driven out of equilibrium by an ultrafast optical pump and then probed with short, delayed X-ray pulses with energy and momentum $({\bf k_i},\omega_i)$ \textemdash resonantly tuned to a specific absorption edge \textemdash which are scattered off the sample with energy and momentum $({\bf k_f},\omega_f)$. The scattered photon energy is resolved by spatially separating different spectral components onto a charge-coupled device (CCD) detector with either a diffraction grating or a crystal analyzer. {\bf b} Time-dependent sequence of the trRIXS process. Adapted with permission from Ref.~\onlinecite{wang2020time} Copyrighted by the American Physical Society.}
    \label{fig:fig1_trRIXS_sketch}
\end{figure}

\begin{figure*}[!ht]
    \centering
    \includegraphics[width=0.9\textwidth]{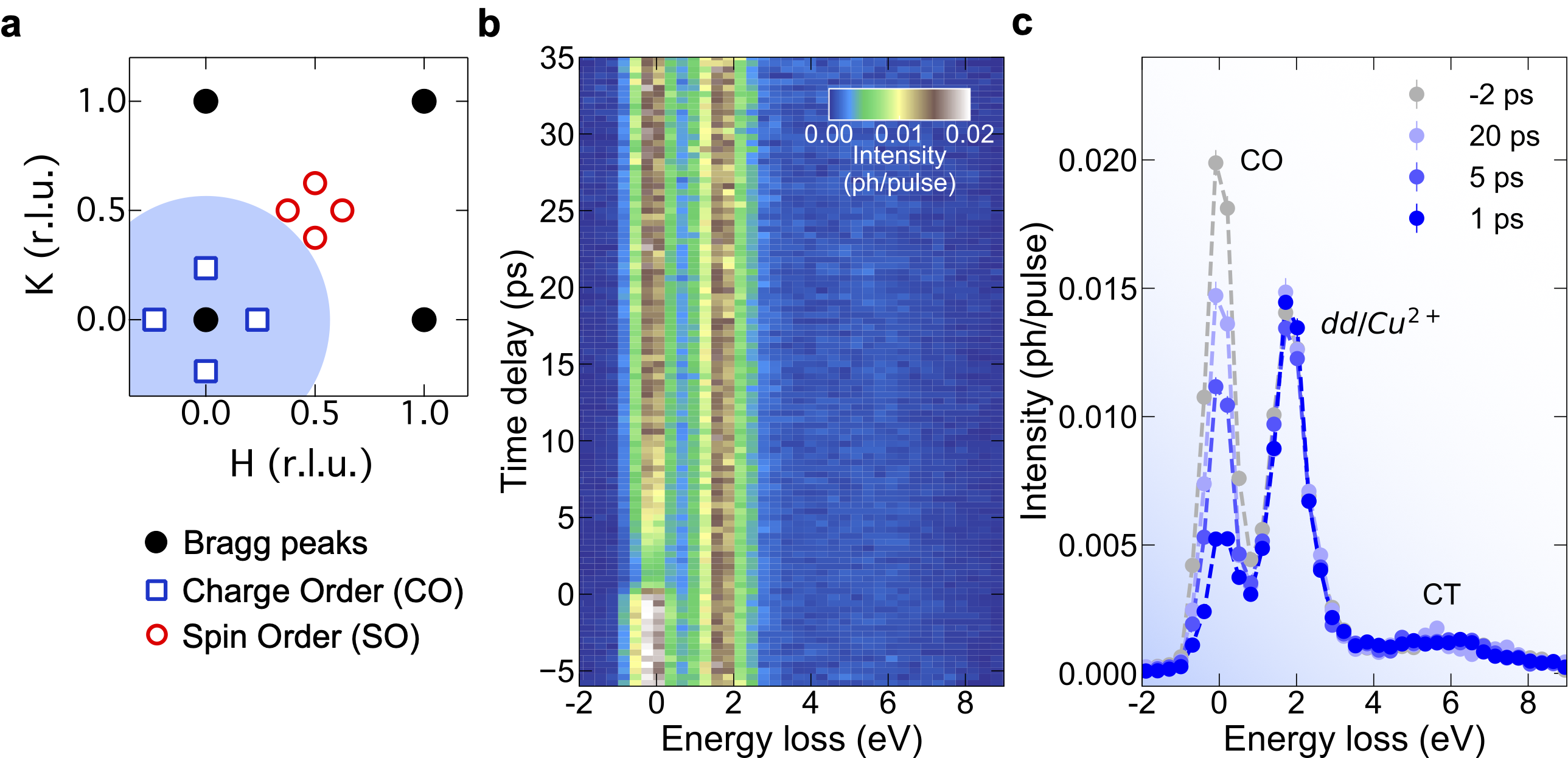}
    \caption{\textbf{Charge order melting in La$_{2-x}$Ba$_x$CuO$_4$.} {\bf a} Reciprocal space map of structural Bragg peaks, charge order (CO) and spin order (SO) peaks in La$_{2-x}$Ba$_x$CuO$_4$. The blue area represents the maximum momentum transfer achievable at the Cu $L$-edge (931 eV). The location of peaks in momentum space is denoted by the Miller indices $(H,K,L)$ and expressed in reciprocal lattice units (r.l.u.) {\bf b} Time-resolved resonant inelastic X-ray scattering (trRIXS) spectra of photoexcited La$_{1.875}$Ba$_{0.125}$CuO$_4$ at $Q_{\rm CO}=(0.236,0,1.5)$ r.l.u. at the Cu $L$-edge for variable pump-probe time delay. {\bf c} trRIXS spectra for a selection of time delays showing a prompt melting of the CO, while higher energy features of the inelastic spectrum, e.g. $dd$ excitations/Cu$^{2+}$ emission and charge transfer (CT) excitations, remain unaffected. Error bars represent Poisson counting uncertainties. Adapted with permission from Ref. \onlinecite{mitrano2019sa}.}
    \label{fig:fig3_mitrano_exp}
\end{figure*}

\par In nonequilibrium experiments, a sample is typically excited by a pump pulse and monitored by a subsequent probe. Interpreting the properties of a photoexcited material, especially when different instabilities are intertwined, requires precise knowledge of how the lattice, band structure, and collective fluctuations respond to the pump. To meet these needs, the ultrafast community developed ultrafast X-ray and electron diffraction for monitoring the crystal lattice~\cite{Siders1999,Fritz2007,SokolowskiTinten2003,Gedik2007} and time- and angle-resolved photoemission (trARPES) for probing the electronic structure\cite{Schmitt2008,Rohwer2011,Smallwood2012}. On the other hand, light-driven collective excitations are commonly investigated with ultrafast optical methods~\cite{Ulbricht2011,Giannetti2016,Nicoletti2016}, which however cannot probe their dispersion in reciprocal space due to the negligible momentum of optical photons. This implies that the microscopic distribution of nonequilibrium fluctuations in quantum materials is largely inaccessible to most existing experimental methods.
\par Time-resolved resonant inelastic X-ray scattering (trRIXS) is a momentum-resolved spectroscopy aimed to interrogate nonequilibrium collective modes, which has been recently enabled by the development of femtosecond X-ray free electron lasers (XFELs) \cite{Bostedt2016,cao2019}.
As shown in Fig.~\ref{fig:fig1_trRIXS_sketch}a, trRIXS probes nonequilibrium dynamics by scattering ultrashort X-ray pulses tuned to a characteristic atomic absorption edge. Once the incoming X-ray photon is absorbed, the pump-excited material transitions to an intermediate state in which a core-level electron is transferred to (or above) the valence orbitals. Within a few fs, the highly unstable intermediate state decays and a valence electron fills the core hole by emitting a X-ray photon (see Fig.~\ref{fig:fig1_trRIXS_sketch}b). The scattered X-rays are then analyzed in both momentum and energy, yielding information about the pump-induced collective dynamics. The resonance condition greatly enhances the RIXS cross section, but the many-body interactions in the intermediate state are what really makes trRIXS sensitive to a wide variety of charge, orbital, and spin excitations of the valence electrons\cite{Ament2011,Haverkort2010}. 

\par In this Perspective, we survey recent experimental and theoretical progress in trRIXS. Then, we outline future research opportunities emerging from these new spectroscopic capabilities.
While trRIXS will have a tremendous cross-disciplinary impact, ranging from chemistry \cite{Wernet2015,Jay2018,Lundberg2019} to condensed matter physics, here we specifically focus on quantum materials' research.

\section{A unique experimental tool}\label{sec:experiment}
The Stanford Linac Coherent Light Source (LCLS) has been at the forefront in developing trRIXS capabilities in both the soft and hard X-ray regime\cite{Bostedt2016,cao2019}. Over the last five years, these developments motivated a variety of experiments focused on probing nonequilibrium correlations in light-driven quantum materials, especially in connection to the problem of high-$T_{\textrm{c}}$ superconductivity.
\par Key challenges in the physics of high-$T_{\textrm{c}}$ superconductors are understanding the relationship between superconductivity and other low-temperature instabilities, as well as devising routes to further enhance $T_{\textrm{c}}$.
In the case of copper oxides, while superconductivity appears upon doping, hole-like carriers also form unidirectional charge (and sometimes spin) order (CO) modulations close to 1/8 doping and at temperatures above the superconducting $T_{\textrm{c}}$, which result in diffraction peaks at a finite momentum {\bf Q}$_{\rm CO}$\cite{Tranquada2004,Abbamonte2005,Ghiringhelli2012,Comin2014,daSilvaNeto2014,Huang2017,Zheng2017}.

\begin{figure*}[!ht]
    \centering
    \includegraphics[width=0.9\textwidth]{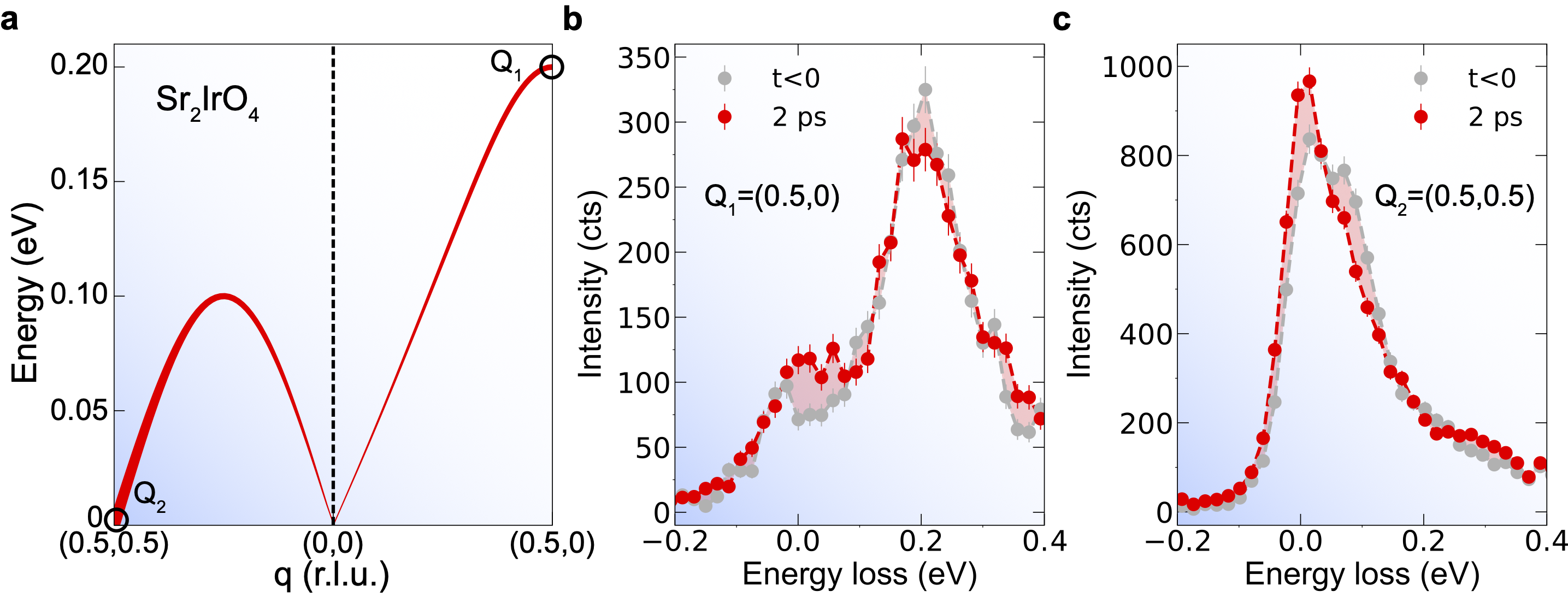}
    \caption{\textbf{Driven magnetic excitations in Sr$_2$IrO$_4$.} {\bf a} Dispersion of the in-plane pseudospin excitation spectrum in Sr$_2$IrO$_4$ as function of momentum transfer $q$. Black circles correspond to momentum transfer points shown in the other two panels. {\bf b}-{\bf c} Ir $L$-edge (11.215 keV) time-resolved resonant inelastic X-ray scattering (trRIXS) snapshots of the pseudospin excitations excited with a 620 meV pump pulse at negative time delay $t$ (grey) and after 2 picoseconds (red)  at momentum transfer {\bf Q}$_1=(0.5,0)$ and {\bf Q}$_2=(0.5,0.5)$ reciprocal lattice units (r.l.u.).  The intensity is measured in counts (cts), while error bars represent Poisson counting uncertainties. The light-induced spectral weight change is highlighted by a red shading. Adapted with author's permission from Ref. \onlinecite{dean2016}}
    \label{fig:fig2_dean_exp}
\end{figure*}

Experimental and theoretical evidence suggest indeed that these two phases interplay and often compete \cite{Ghiringhelli2012,Chang2012,Jiang2019}. Furthermore, ultrafast optical pulses have been found to enhance superconductivity while melting charge order correlations \cite{Hu2014,Kaiser2014,Fausti2011,Nicoletti2014,Forst2014a,Forst2014b}, and vice versa\cite{Wandel2020}.
Understanding how light affects the balance between these two phases and their collective dynamics with the aim of further optimizing superconductivity requires measuring the transient inelastic charge spectrum. 
\par To this end, a recent trRIXS experiment at the Cu $L$-edge investigated the light-induced charge order dynamics of the prototypical stripe-ordered cuprate La$_{2-x}$Ba$_x$CuO$_4$ (see Fig. \ref{fig:fig3_mitrano_exp}{\bf a}) \cite{mitrano2019sa,mitrano2019prb}. trRIXS spectra clearly show that 1.55-eV photons, which transiently enhance interlayer superconducting tunneling \cite{Nicoletti2014,Cremin2019}, also deplete the quasielastic charge order peak at {\bf Q}$_{\rm CO}$ (see Fig. \ref{fig:fig3_mitrano_exp}{\bf b-c}). Unlike conventional charge density waves \cite{Huber2014}, the CO is found to undergo a sudden light-induced sliding motion\cite{mitrano2019prb} and exhibit an exponential recovery dominated by yet unobserved diffusive fluctuations at the sub-meV scale\cite{mitrano2019sa}. By applying a similar approach to other copper oxides such as YBa$_2$Cu$_3$O$_{7-\delta}$ and Nd$_{1+x}$Ba$_{2-x}$Cu$_3$O$_{7-\delta}$\cite{Arpaia2019}, we expect trRIXS to provide new insights about the broader dynamical relationship between charge order and superconductivity.  

\par Aside from charge dynamics, spin fluctuations (particularly near the antiferromagnetic wavevector {\bf Q}$_{\textrm{AF}}=(0.5,0.5)$ r.l.u.) are believed to participate to the superconducting pairing \cite{scalapino1986d,gros1987superconducting,kotliar1988superexchange,tsuei2000pairing,Scalapino2012,maier2016pairing}. Thus, their optical excitation is a promising route to manipulate nonequilibrium superconductivity. Being sensitive to spin degrees of freedom through the intermediate state, trRIXS is the only available method for measuring the transient magnetic excitation spectrum as a function of momentum.
Pioneering experiments provided a first glimpse of light-induced spin dynamics in a Mott insulator \cite{dean2016,mazzone2020}. 
Unlike Cu $L$-edge X-rays (see Fig.\ref{fig:fig3_mitrano_exp}a), photons at the $L$-edge of 5$d$ transition metals carry enough momentum to fully map magnetic fluctuations throughout the Brillouin Zone (see Fig.\ref{fig:fig2_dean_exp}a). Among the 5$d$ materials, iridates (Sr$_{n+1}$Ir$_{n}$O$_{3n+1}$) are particularly interesting analogues of copper oxides. These compounds may give rise to unconventional superconductivity upon doping \cite{Wang2011,Kim2014}, with pseudospin fluctuations (due to spin-orbit coupling) playing the same role as spins in cuprates.
Driving onsite orbital excitations with infrared pump pulses resulted in a significant spectral weight reshaping of pseudospin excitations at the high symmetry points {\bf Q}$_1=(0.5,0)$ and {\bf Q}$_2=(0.5,0.5)$ reciprocal lattice units (r.l.u.) (see Fig.\ref{fig:fig2_dean_exp}b-c) \cite{dean2016,mazzone2020}. These first snapshots of light-stimulated magnetic excitations at the Ir $L$-edge show that optical pump pulses with a near-zero momentum transfer have profound influence on the finite-momentum spin dynamics of quantum materials throughout the Brillouin zone.
\par These experiments, along with further studies of orbital excitations \cite{parchenko2020,DellAngela2016}, demonstrate how trRIXS provides simultaneous access to charge, spin, and orbital degrees of freedom far from equilibrium. Its sensitivity to collective fluctuations at large momenta fills a long-standing gap in time-resolved experiments and makes trRIXS a unique tool to advance our microscopic understanding of light-driven phenomena in quantum materials.

\section{An evolving theoretical framework}

A fundamental need for current and future trRIXS research lies in calculating the cross-section and predicting the many-body response of photoexcited quantum materials. Unlike other spectroscopies (e.g. trARPES, and non-resonant light/x-ray scattering)\cite{freericks2009theoretical, wang2017, freericks2018nonresonant, wang2018theory}, the RIXS process involves a four-time correlation function, due to the presence of the resonant intermediate state \cite{chen2019, wang2020time}. While this complexity entails modeling and numerical challenges, it is precisely the intermediate state dynamics which makes RIXS sensitive to a wide variety of collective excitations. Moreover, these excitations encode high-order correlations beyond the linear response, and thus play a crucial role in emergent phenomena with strong quantum fluctuations.

\par If the intermediate state is assumed to last for a negligibly short time [ultrashort core-hole lifetime (UCL) approximation], the dominant contribution to the equilibrium RIXS spectrum comes from the dynamical structure factors. \cite{VanDenBrink2007, ament2007ultrashort, Ament2011,jia2016using}.
For this reason, early theories of trRIXS focused on calculating charge and spin structure factors in correlated electron models driven out of equilibrium\cite{wang2017,paeckel2019}. For example, an early simulation of a one-dimensional Mott insulator excited by a realistic below-gap pump pulse found evidence of light-induced Floquet replicas in both the spin and charge response. Different from an ideal Floquet picture, these light-engineered excitations persisted and evolved after the pump pulse \cite{wang2017}, thus indicating that light pulses are conducive to emergent Floquet dynamics at finite momentum.

\begin{figure*}
    \centering
    \includegraphics[width=0.9\textwidth]{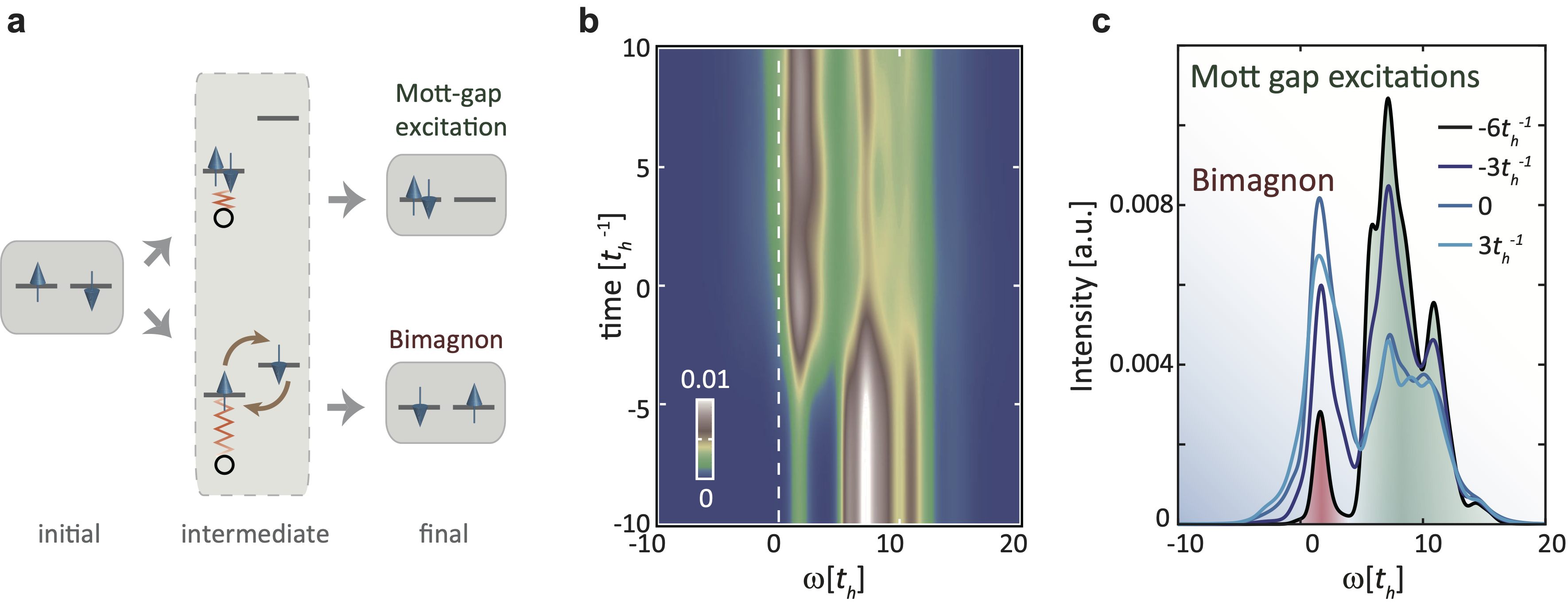}
    \caption{\textbf{Theoretical simulation of the time-resolved resonant inelastic X-ray scattering (trRIXS) spectrum in a pumped Mott insulator.} {\bf a} Dominant excitations in the $K$-edge (indirect) trRIXS spectrum of the single-band Hubbard model in two dimensions. The open circle represents the core hole, while the red zig-zag line is its Coulomb interaction with the valence electron spins (up down arrows). {\bf b} Evolution of the trRIXS spectrum obtained through exact diagonalization for momentum transfer $\mathbf{Q}=(1/2,1/6)$ reciprocal lattice units (r.l.u.) and incident photon energy tuned to the maximum of the X-ray absorption spectrum. $\omega$ denotes the energy loss axis measured in units of hopping energy ($t_{\rm h}$). {\bf c} Selected trRIXS spectra at different times before and after the pump. Bimagnon features are located at $\hbar\omega\sim 3J$, while the Mott gap excitation continuum is peaked at $\hbar\omega\sim U=8t_h$. $U$ is the onsite Coulomb repulsion, while $J=4t_{\rm h}^2/U$ is the exchange energy.  Adapted with permission from Ref. \onlinecite{wang2020time}. Copyrighted by the American Physical Society.} 
    \label{fig:fig4_theory}
\end{figure*}

\par While these results constitute an exciting starting point, the UCL approximation ignores physical processes faster than the timescale of the core-hole lifetime (longer than 2 fs at the Cu $L$ edge\cite{Ament2011,rossi2019experimental}) as well as high-order correlations\cite{tohyama2018spectral}. Therefore, a simple calculation of the structure factors may not capture the full trRIXS spectrum and miss nonlinear spectral features such as bimagnons and $dd$ excitations.
In order to compute the full trRIXS cross-section, one needs to explicitly model the resonant probe process and account for the finite intermediate-state lifetime, leading to a computational complexity $O(N_{\rm t}^4)$, where $N_{\rm t}$ is the number of evolution steps\,\cite{chen2019, wang2020time}. 
\par Figure \ref{fig:fig4_theory} shows the numerically-calculated trRIXS spectrum of a two-dimensional Hubbard model probed through an indirect scattering process (e.g. Cu $K$-edge) and explicitly accounting for intermediate state effects. Here, excitations are generated through interactions between the valence electrons and the core hole during the intermediate state\,\cite{wang2020time}. Unlike the structure factors obtained within the UCL approximation, the trRIXS cross-section contains high-order excitations including the bimagnon at energy $\sim 3J$ ($J=4t_{\rm h}^2/U$ is the spin-exchange interaction, $t_{\rm h}$ is the hopping amplitude, and $U$ is the onsite Coulomb repulsion), in addition to Mott-gap excitations.
After being driven by a pump resonant with the Mott gap, spectral weight gets transferred from the Mott peak to in-gap excitations and bimagnons. In this case, the visibility of Floquet replicas of collective excitations is reduced when compared to Ref.~\onlinecite{wang2017,chen2019}, mainly due to the shorter Floquet state lifetime in the presence of electron-electron interactions.
The predicted in-gap spectral weight transfer resembles the quasielastic scattering intensity enhancement observed in the iridate $L$-edge trRIXS spectra (see Fig.~\ref{fig:fig2_dean_exp}), and could be measured in future $K$-edge experiments on gapped correlated materials. 
\par Intermediate state effects are not only required to capture richer physics in trRIXS simulations, but will also guide the interpretation of new types of inelastic scattering experiments. RIXS spectra collected while slightly detuning the incident X-ray pulse energy away from resonance can be used to selectively enhance the intensity of excitations involving specific intermediate states \cite{wang2020time, chen2020observing} and to extract information about momentum-resolved electron-phonon coupling of well-isolated modes \cite{ament2011determining, rossi2019experimental,Geondzhian2020}. Detuning experiments, in tandem with microscopic calculations, will be particularly impactful in interrogating the electron-phonon coupling in photoexcited charge density waves \cite{Huber2014,Lee2012,Wandel2020}, light-induced ferroelectricity \cite{Nova2019} and superconductivity \cite{Mankowsky2014}.
\par In summary, advances in trRIXS experiments are accompanied by a steadily evolving theoretical framework aimed at understanding the observed light-induced dynamics. Current computational capabilites and state-of-the-art algorithms allow to perform accurate and predictive nonequilibrium simulations of the trRIXS spectrum in photoexcited correlated electron systems. In the future, we expect trRIXS theory to play an ever important role not just in interpreting time-resolved scattering experiments, but in leading the field towards new discoveries.

\section{A new scientific opportunity}

\begin{figure*}
    \centering
    \includegraphics[width=\textwidth]{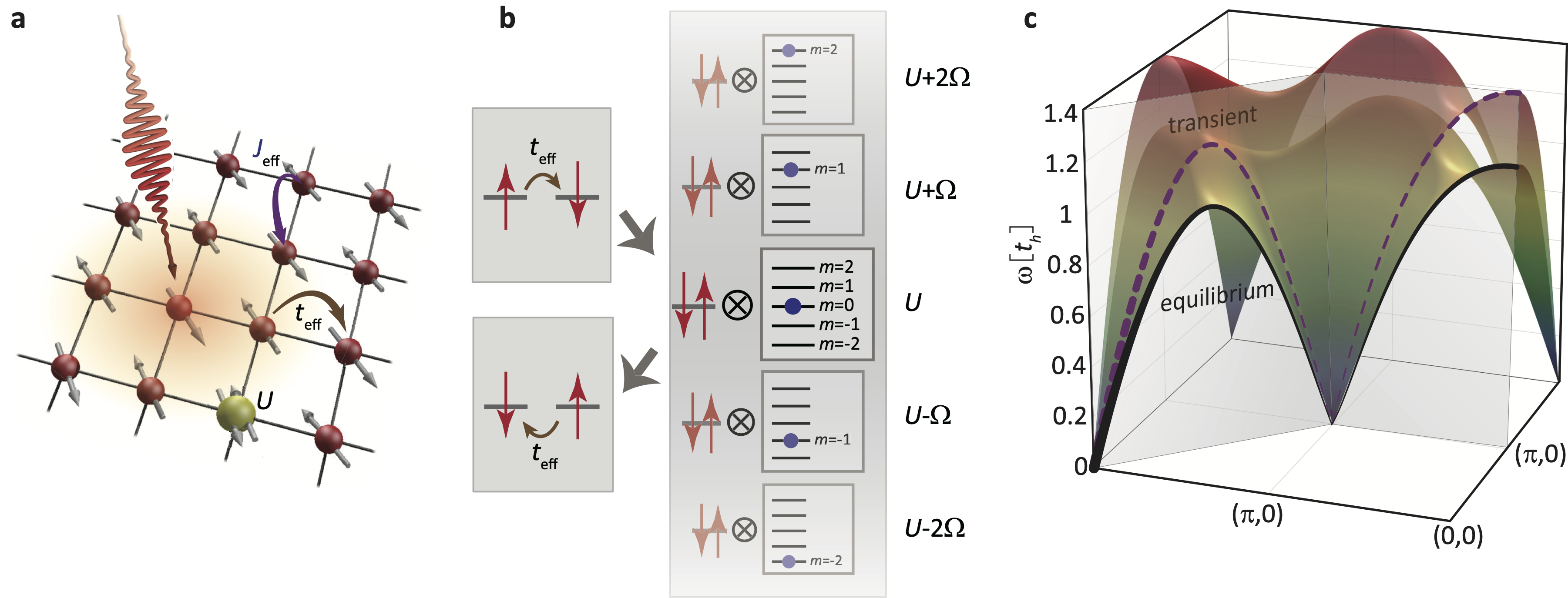}
    \caption{\textbf{Ultrafast manipulation of finite-momentum spin waves.} {\bf a} Ultrafast pulses can be used to directly manipulate magnetic excitations in quantum materials through a direct modulation of the effective exchange interactions $J_{\textrm{eff}}$. {\bf b} A possible mechanism consists in dressing the intermediate doubly-occupied state with the pump field, which is lifted (lowered) in energy by an amount proportional to the number of virtually absorbed (emitted) pump photons. $U$ is the onsite Coulomb repulsion, $t_{\textrm{eff}}$ the effective hopping amplitude, $\Omega$ the pump photon energy and $m$ the Floquet index {\bf c} As the exchange interactions are modulated, the spin wave dispersion is expected to shift in energy ($\omega$) across the entire Brillouin zone. A cross-section cut (grey shaded planes) along two high-symmetry directions is included for clarity.} 
    \label{fig:fig5_magnon}
\end{figure*}
As shown in previous sections, trRIXS is rapidly growing into a major spectroscopy of light-driven quantum materials, and its future developments are critically tied to the pace of technological advances at XFEL facilities.
Owing to the small inelastic cross sections, trRIXS will benefit from a dramatic increase in the average XFEL spectral brightness from the current LCLS $10^{20}$ to $10^{24}-10^{25}$ photons s$^{-1}$ mm$^{-2}$ mrad$^{-2}$ (0.1$\%$ bandwidth)$^{-1}$ at 1 keV\cite{Bostedt2016,Dunne2019}. This brightness enhancement will be mainly achieved through increased repetition rates, e.g. at both the LCLS-II and the European XFEL, and will lead to order-of-magnitude improvements of the signal-to-noise ratio.
Furthermore, higher energy resolution (especially in the soft X-ray regime) will enable observing low-energy collective fluctuations, such as phonons and spin waves. At the time of this writing, large spectrometers are in construction at both the LCLS-II (NEH2.2/q-RIXS instrument)\cite{Dunne2019} and the European XFEL (hRIXS instrument). The target resolving power at 1 keV would be of order $3.0\cdot10^4$ ($\sim0.03$ eV resolution), thus implying a 20x improvement with respect to the data shown in Fig. \ref{fig:fig3_mitrano_exp} \cite{Dunne2018}. Other experimental endstations with different features are being developed at the Trieste Free Electron laser Radiation for Multidisciplinary Investigations (FERMI/Italy), the Pohang Accelerator Laboratory X-ray Free Electron Laser (PAL-XFEL/S. Korea), and the Switzerland’s X-ray free-electron laser (SwissFEL). 
These enhanced capabilities open up a new frontier for the investigation of nonequilibrium spin, charge and orbital dynamics in strongly correlated and topologically non-trivial materials.
\par A tantalizing application of trRIXS is the study of spin excitations in light-driven quantum materials. Multiple theoretical studies have proposed that spin-exchange interactions can be controlled by renormalizing the effective Hamiltonian interactions through a periodic drive (see Fig.~\ref{fig:fig5_magnon}\textbf{a}), in the spirit of the so-called ``Floquet-engineering"\,\cite{Mentink2014,Mentink2015,Chaudhary2019,Walldorf2019}. In this idealized scheme, photons dress the intermediate electronic states of the exchange (or superexchange) process, leading to an effective energy scale $J_{\rm eff}$ (see Fig.~\ref{fig:fig5_magnon}\textbf{b}). Although the parameter renormalization is rigorous only for an infinitely-long periodic pump, it could also be achieved with an ultrashort laser pulse \cite{wang2018theory}. Alternatively, the spin excitations could be dynamically altered through other protocols, e.g. by distorting the lattice via the nonlinear phonon coupling \cite{Forst2011,Subedi2014}.
Once $J_{\rm eff}$ is modified by the pump, trRIXS measurements will interrogate changes in the dispersion, linewidth and spectral weight of the spin fluctuation spectrum (see Fig.~\ref{fig:fig5_magnon}\textbf{c}). 
A measurement of the dispersion allows us to disentangle renormalization effects of multiple coexisting exchange interactions\cite{Peng2017}, which cannot be discerned from the bimagnon peak in the Raman spectrum.
Furthermore, a lineshape analysis throughout the Brillouin zone enables exploring excitations beyond magnons, e.g. incoherent fluctuations in geometrically frustrated lattices \cite{sandilands2015scattering,chun2015direct,gretarsson2013magnetic,nasu2016fermionic}. 
Since the experimental energy resolution will be of order $\sim0.03$\,eV in the near future\cite{Dunne2018,Dunne2019}, this technique will be particularly effective for studying materials with relatively large exchange scales, such as cuprates, iridates and certain nickelates, where the spin excitations disperse over energies larger than 0.1 eV.

\begin{figure*}[!t]
    \centering
    \includegraphics[width=0.9\textwidth]{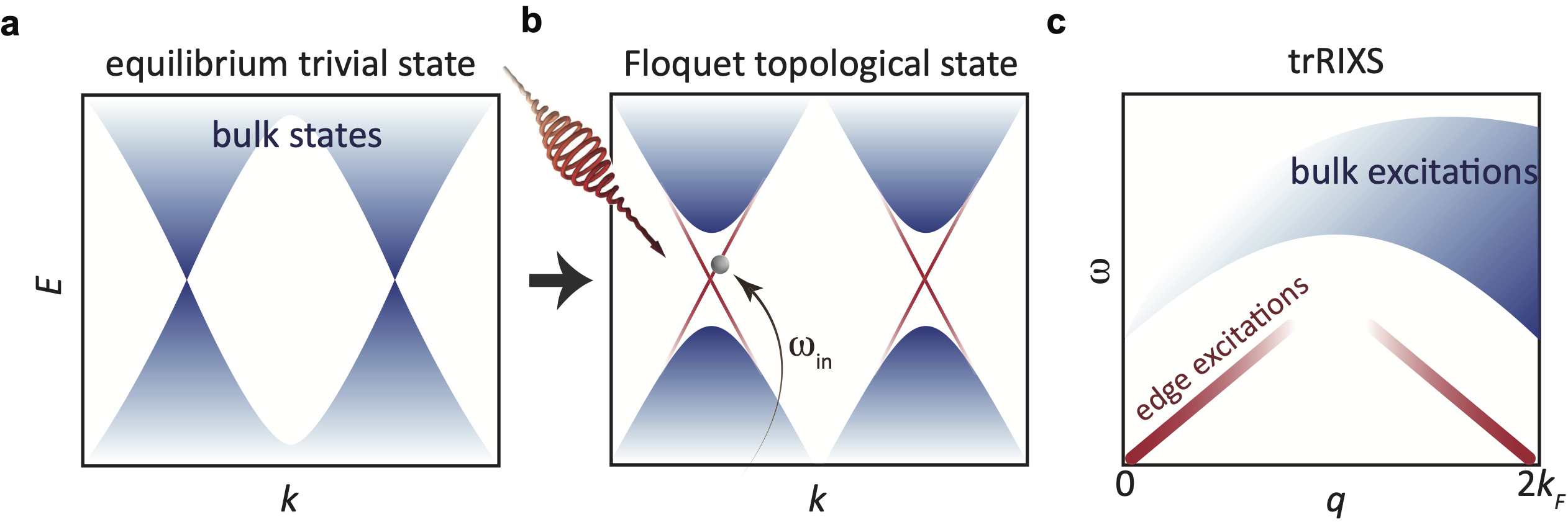}
    \caption{\textbf{Generation and detection of topological collective modes.} {\bf a} Circularly-polarized pump pulses can drive bulk Dirac carriers (blue) and {\bf b} open a gap at the Dirac points by breaking time-reversal symmetry. $E$ and $k$ denote energy and momentum of the band structure, respectively, while $\omega_{\rm in}$ stands for the incoming X-ray photon energy. In addition to the light-induced Floquet bandgap, the system exhibits chiral edge states (red) which give rise to dispersing collective modes in the time-resolved resonant inelastic X-ray scattering (trRIXS) spectrum {\bf c}. These collective modes are linearly dispersing and only present at overlap with the pump excitation, unlike the particle-hole continuum due to the bulk states. $\omega$ and $q$ indicate energy loss and momentum transfer, while $k_{\rm F}$ represents the Fermi wavevector.} 
    \label{fig:fig6_topology}
\end{figure*}

\par In addition to spin fluctuations, trRIXS will play as well a pivotal role as a probe of ultrafast charge and orbital dynamics.
We envision here two promising research directions involving the charge sector of low-dimensional quantum materials, namely the time-resolved investigation of fractionalized excitations, and the search for new photoinduced condensation phenomena.
In one dimension, electrons cannot propagate freely, but instead displace their neighbors due to electron-electron interactions. This leads to a breakdown (``fractionalization") of the electron in a variety of collective excitations propagating with different velocities \cite{Giamarchi2004}. trRIXS provides an opportunity to study these fundamental phenomena in real time by exciting a transient particle-hole plasma with a high photon energy pump and directly observing the time-dependent behavior of the collective modes contained in the transient RIXS spectrum. First experiments could focus on fractionalization in Sr$_2$CuO$_3$ \cite{Schlappa2012} or CaCu$_2$O$_3$ \cite{Bisogni2015}, and on manipulating their excitations by dynamically tuning the balance between spin-orbit coupling and crystal field\cite{chen2015fractionalization}. Similarly, trRIXS experiments in photoexcited Mott insulators could reveal the fingerprints of new condensation phenomena, such as $\eta$-pairing\cite{Yang1989,Zhang1990} and dynamical $p$-wave superconductivity\cite{werner2018enhanced}. The $\eta$-paired phase is a superfluid of doubly-occupied electronic states carrying finite momentum ${\bf Q}_{\eta}$ and arising from a broken SU(2) symmetry of the Hubbard Hamiltonian. Since the $\eta$-pairing is an eigenstate, but not necessarily a ground state, pump light pulses open the possibility of stabilizing this yet unobserved phase\cite{Sota2016,Kaneko2019,Cook2020,Ryo2020,Peronaci2020}. A periodic pump is theoretically predicted to enhance pairing correlations and establish true off-diagonal long-range order by dynamically renormalizing the onsite Coulomb repulsion\cite{Peronaci2020}. trRIXS would then search for $\eta$-pairing signatures in the finite-momentum pairing susceptibility\cite{Suzuki2019}, namely a divergent quasi-elastic structure factor at ${\bf Q}_{\eta}$ and a triplet of collective modes at energies $\hbar\omega=0,\pm(U-2\mu)$ ($\mu$ being the chemical potential)\cite{Zhang1990}.
Beyond $\eta$-pairing, resonantly driving orbital degrees of freedom in doped Mott insulators has also been proposed as a route to dynamically stabilize $p$-wave superconductivity\cite{werner2018enhanced}.

\par The search for new light-driven phenomena in materials dominated by local electronic correlations also calls for the development of more advanced spectroscopic methods. One such approach is trRIXS interferometry. Thanks to the local nature of the intermediate core holes and the intrinsic coherence of the scattering process, the RIXS signal can indeed exhibit interference among different intermediate states \cite{Ma1995}. In the dimerized spin-orbit coupled insulator Ba$_3$CeIr$_2$O$_9$ \cite{Revelli2019}, the intermediate state involves a coherent superposition of a single core hole on either of the two atoms in the dimer. This leads to a $\cos^2({\bf Q}\cdot{\bf d}/2)$ modulation of the RIXS intensity in momentum space (${\bf d}$ being the intradimer distance). Importantly, the interference pattern varies in amplitude and phase depending on the symmetry of the excited-state wavefunction \cite{Revelli2019}, thus providing an interferometric measurement of the local atomic orbitals. By sampling multiple interference fringes in the hard X-ray regime (e.g. Ir $L$-edge, and Cu $K$-edges), it would be possible to exquisitely resolve transient changes to the local electronic structure down to the picometer level with energy selectivity. Not just limited to dimerized compounds, this technique could likewise be applied to study nonequilibrium dynamics of confined electrons along one or two directions, such as in ladder compounds \cite{Abbamonte2004} or layered cuprates \cite{Mankowsky2014} and nickelates \cite{Zhang2016,Zhang2017}. 
\par In a very different context, confined electronic motion is also a defining property of edge states in light-induced topological phases, which could be revealed by trRIXS experiments\cite{chen2020observing}. Circularly polarized laser pulses have been shown to break time-reversal symmetry and induce transient states with non-trivial Chern numbers\cite{haldane1988model,claassen2016all,Hubener2017,McIver2019}. An intriguing application of this experimental approach is the creation of tunable Floquet topological insulators (FTIs), in which topology can be manipulated by varying pump amplitude, energy, and polarization\cite{kitagawa2011transport, gu2011floquet, usaj2014irradiated, torres2014multiterminal, dehghani2015out}.  
As shown in Fig.~\ref{fig:fig6_topology}a, a possible route for creating a FTI in two dimensions starts from a material with bulk massless Dirac fermions. The circularly-polarized pump induces a gap opening at the Dirac point and chiral edge modes at the sample boundary, which disperse across the light-induced bandgap (see Fig.~\ref{fig:fig6_topology}b). Detecting these edge states entails probing either the transient band structure with trARPES, or their collective modes in the dynamic structure factor through trRIXS, depending on the experimental constraints. However, trRIXS offers a crucial advantage. By resonantly tuning the incident photon energy to transitions from core states into the light-induced bandgap, this technique can boost the visibility of the topological edge states over the bulk signal, and hence distinguish their dispersion from other bulk collective modes\cite{chen2020observing} (see Fig.~\ref{fig:fig6_topology}c). Future developments in nano-trRIXS may enable the direct spatial imaging of edge state dynamics and, thus, further increase their visibility over bulk excitations (although these experiments would require a special handling of X-ray irradiation effects).
Finally, an additional area of interest (particularly for hard X-ray trRIXS) is the light-control of candidate topological superconductors under high pressures \cite{zhang2011pressure,kirshenbaum2013pressure, zhou2016pressure,he2016pressure}, which are inaccessible to photoemission experiments.
\vspace{2mm}
\par This short, and by no means complete, array of examples underscores how trRIXS measurements, alongside new theoretical methods, will play an essential role in detecting and understanding new dynamic phenomena in light-controlled quantum materials. Increased XFEL beamtime availability, X-ray brightness and spectrometer performance will enable more sophisticated trRIXS experiments with higher energy resolution and polarization control.
The possibilities opened by these advances are difficult to grasp, but are certainly positioning trRIXS on the leading edge of a decade of discovery.

\section*{Acknowledgements}

We acknowledge E. Baldini, M. Buzzi, R. Comin, G. Coslovich, M. P. M. Dean, J. Freericks, A. A. Husain, D. Nicoletti, A. H. Reid, and K. Wohlfeld for valuable discussions. We also thank M. P. M. Dean for providing the original data from Ref.~\onlinecite{dean2016}. The experiments reported in Fig. \ref{fig:fig3_mitrano_exp} were supported by the U.S. Department of Energy, Office of Basic Energy Sciences grant no. DE-FG02-06ER46285. Use of the Linac Coherent Light Source (LCLS), SLAC National Accelerator Laboratory, is supported by the U.S. Department of Energy, Office of Science, Office of Basic Energy Sciences under Contract No. DE-AC02-76SF00515. The calculation reported in Fig.~\ref{fig:fig4_theory} used resources of the National Energy Research Scientific Computing Center (NERSC), a U.S. Department of Energy Office of Science User Facility operated under Contract No. DE-AC02-05CH11231. 

\section*{Author contributions} Both authors, M. M. and Y. W., wrote the manuscript. 

\section*{Competing interests} The authors declare no competing interests.

\bibliography{reviewRef}
\end{document}